\documentclass[useAMS,usenatbib]{mn2e}
\usepackage{epsfig}
\usepackage{natbib}

\title[The radio counterpart of HESS\,J0632+057]{The radio counterpart of the likely TeV binary HESS\,J0632+057}
\author[Skilton et al.]{J.L. Skilton$^{1}$\thanks{E-mail:J.L.Skilton03@leeds.ac.uk},
 M. Pandey-Pommier$^{2}$, J.A. Hinton$^{1}$, C.C. Cheung$^{3}$,
\newauthor F.A. Aharonian$^{4}$, J. Brucker$^{5}$, G. Dubus$^{6}$, A. Fiasson$^{7,8}$, S. Funk$^{9}$,
\newauthor Y. Gallant$^{7}$, A. Marcowith$^{7}$, O. Reimer$^{9,10}$\\
$^{1}$School of Physics and Astronomy, University of Leeds, Leeds, LS2 9JT, UK\\
$^{2}$Leiden Observatory, P.O. Box 9513, 2300 RA, Leiden, The Netherlands \\
$^{3}$NASA Goddard Space Flight Center, Astrophysics Science Division, Code 661, Greenbelt, MD, 20771, USA \\
$^{4}$Dublin Institute for Advanced Studies, 5 Merrion Square, Dublin 2,
Ireland\\
$^{5}$Universit\"at Erlangen-N\"urnberg, Physikalisches Institut, Erwin-Rommel-Str. 1,
D 91058 Erlangen, Germany\\
$^{6}$Laboratoire d'Astrophysique de Grenoble, INSU/CNRS, Universit\'e Joseph Fourier, BP
53, F-38041 Grenoble Cedex 9, France\\
$^{7}$Laboratoire de Physique Th\'eorique et Astroparticules, CNRS/IN2P3,
Universit\'e Montpellier II, CC 70, Place Eug\`ene Bataillon,\\
F-34095 Montpellier Cedex 5, France\\
$^{8}$Laboratoire d'Annecy-le-vieux de Physique des Particules, CNRS/IN2P3, 9
chemin de Bellevue, B.P. 110 F-74941,\\ Annecy-le-Vieux Cedex, France\\
$^{9}$Kavli Institute for Particle Astrophysics and Cosmology, SLAC, 2575 Sand Hill Road, Menlo-Park, CA-94025, USA\\
$^{10}$Institute for Astro and Particle Physics, Innsbruck University, A-6020 Innsbruck, Austria }
\begin{document}

\date{}

\pagerange{\pageref{firstpage}--\pageref{lastpage}} \pubyear{2002}

\maketitle

\label{firstpage}

\begin{abstract}
The few known $\gamma$-ray binary systems are all associated with
variable radio and X-ray emission. The TeV source HESS J0632$+$057,
apparently associated with the Be star MWC\,148, is plausibly
a new member of this class. Following the identification of a
variable X-ray counterpart to the TeV source we conducted GMRT and VLA
observations in June-September 2008 to search for the radio
counterpart of this object. A point-like radio source at the
position of the star is detected in both 1280\,MHz GMRT and 5\,GHz
VLA observations, with an average spectral index, $\alpha$, of $\sim$0.6.
 In the VLA data there is significant flux variability
on $\sim$month timescales around the mean flux density of $\approx$0.3~mJy.
These radio properties
(and the overall spectral energy distribution) are consistent with an
interpretation of HESS J0632$+$057 as a lower power analogue of the
established $\gamma$-ray binary systems.

\end{abstract}


\section{Introduction}

\begin{figure*}
\centering
\epsfig{file=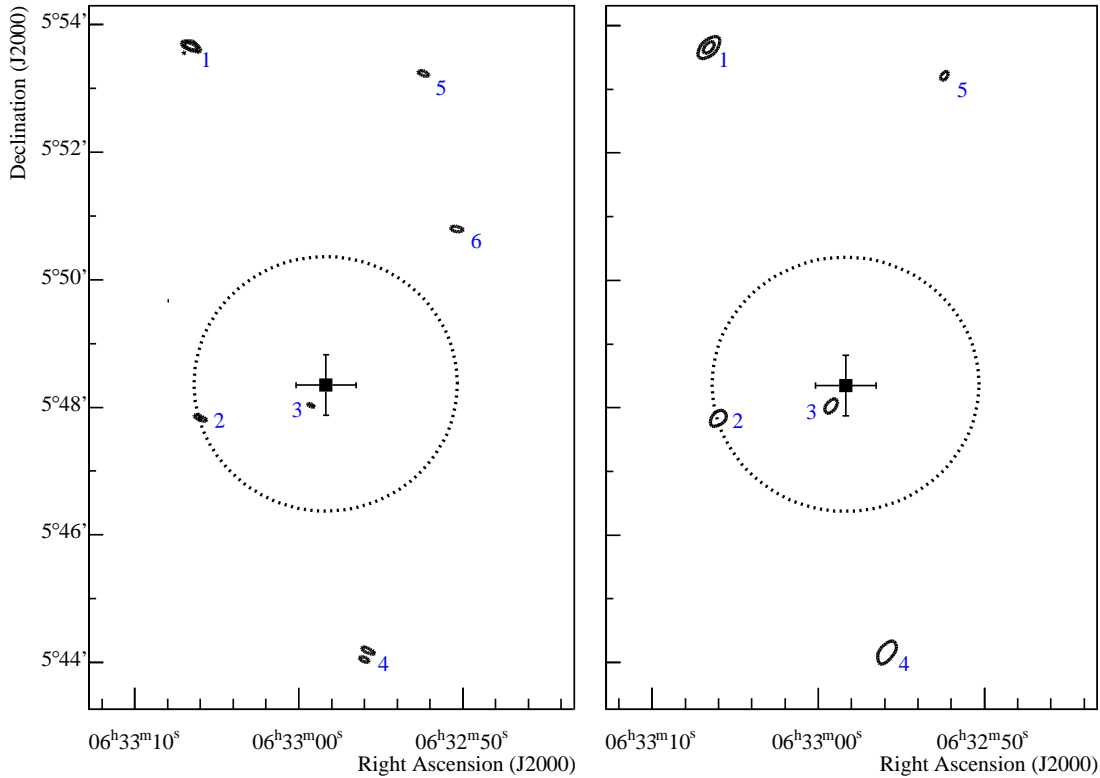, width=0.85\linewidth, clip=} \\
\caption{ Images of the region surrounding HESS\,J0632$+$057 (shown as
  a black square with positional error bars) obtained with the GMRT at 1.28\,GHz (left panel)
and the VLA in D-configuration at 5\,GHz (right panel). Five point-like
sources are detected in both images (note that the increased angular resolution
of the GMRT observations show that source \#4 is in fact a double source). Only
one source (source \#3) lies within the rms size limit of the TeV emission from
HESS\,J0632+057 (dashed black circle). The sources are labelled in order of decreasing R.A.
The flux density contours are at 0.4\,mJy/beam and 2.0\,mJy/beam in the GMRT map (beamsize 2'') and at 
0.2\,mJy/beam and 1.0\,mJy/beam in the VLA map (beamsize 14'').}
\label{fig:image}
\end{figure*}

There are three firmly established $>$GeV emitting binaries;
 PSR\,B1259$-$63 \citep{HESS:1259}, LS\,5039 \citep{HESS:5039} and LS\,I\,+61\,303
\citep{MAGIC:LSI}, all systems composed of a compact object and a high-mass star.
LS\,5039 harbours an O6.5V type star in a 3.9 day orbit, the massive star in the other
systems is of Be type. 
PSR\,B1259$-$63 is the only one of these systems in
which the nature of the compact companion (a 48~ms period radio pulsar in a 3.4~year orbit around the star) has been identified.
All three systems display similarities in their radio
to TeV $\gamma$-ray spectral energy distributions (SEDs) including a hard X-ray spectrum
and a softer TeV spectrum, radio emission above 1\,GHz and %%% $\sim$0.5\%
variability in all bands.
There is also evidence for a single TeV flare from the luminous X-ray binary
Cygnus X-1~\citep{MAGIC:cygnusX}, 
a system of a black hole and an O9.7 Iab star
in a 5.6 day circular orbit. 

The point-like TeV source HESS\,J0632+057~\citep{0632:HESS}, 
has been suggested as a new member of this class of objects. This suggestion was based on 
associations with the massive star MWC\,148 (HD\,259440) and the unidentified ROSAT and EGRET sources
1RXS\,J063258.3+054857 and 3EG\,J0634+0521, and would imply that MWC\,148 has a 
hidden compact companion. X-ray observations of this source with XMM-Newton in 2007 revealed
a point-like, hard spectrum ($\Gamma$$\approx$1.26), X-ray source at the
position of MWC\,148 \citep{0632:XMM}. Significant variability was detected 
on timescales of a few hours.
The association of this X-ray source with the TeV source (and both sources
with MWC\,148) seems highly likely and the spectral properties of the X-ray/$\gamma$-ray source are 
consistent with the known $\gamma$-ray binaries. Very recently the VERITAS
collaboration have published flux upper limits which imply variability in the TeV emission of this object \citep{VERITAS:0632}.

%% This could be reduced a bit I think...
Historical studies of MWC\,148
have determined the star to be of spectral type B0pe \citep{Morgan55}. 
Be stars are classified by the presence of Balmer emission lines.
%%Unfortunately, this broad classification includes many
%%other objects beside classical Be stars. 
MWC\,148 was rejected as a Herbig AeBe
star due to a non-detection during the IRAS survey \citep{The94}
and thus is believed to represent one of the `classical'
Be stars. These stars are fast rotators, thought to be
spinning at 50-90\% of their critical velocity (note \cite{Gutierrez07} give $v$\,sin\,$i$
of MWC\,148 to be 430\,km\,s$^{-1}$) with a large stellar wind and a
%JAH> and what is the critical velocity?
high mass loss rate. The flattening of
the circumstellar envelope produces a global polarisation of the envelope
 of a few percent. The most recent measured values of the polarisation of
 MWC\,148 are 3-4\% \citep{Yudin98}. The star also exhibits a significant infrared
excess \citep[see 2MASS,][]{2MASS:new}
consistent with the existence of an extended envelope. %%?
The source lies between the Monoceros Loop supernova remnant and the star forming regions of the
Rosetta Nebula ($d \approx$1.4\,kpc).
% and is thought to be associated with the stellar cluster NGC 2247 \citep{Kilkenny85}.
A compatible distance estimate of $\approx$1.5\,kpc is calculated using the apparent visual magnitude
of MWC\,148, $M_V$ = 9.1 \citep{Hog00}, as described in \cite{0632:XMM}.
No variability was detected from the star during the ASAS-3 optical survey \citep{Gutierrez07}.
Bp stars are defined by their unusual surface abundances and strong \citep[up to kG, see e.g.][]{Borra79}
surface magnetic fields \citep{Phillips88}.
The confinement of the stellar wind in these strong fields is thought to lead to strong shock heating and
variable X-ray emission. \cite{Townsend07} speculate 
that such a process may be capable of accelerating particles up to TeV energies.

All of the known binaries have been detected at radio wavelengths at
flux levels of 10--100\,mJy \citep[see][and references
therein]{Dubus06}. The radio spectra of LS\,5039 and LS\,I\,+61\,303
have spectral indices close to $\alpha$ = 0.5 (where $F_{\nu} \propto \nu^{-\alpha}$).
\textbf{In LS\,I\,+61\,303 the radio flux is modulated with the orbital period, whereas in LS\,5039, the radio emission (unlike the X-ray and TeV emission) is not \citep[see e.g.][]{Clark01}.}
 No targeted radio observations of
MWC\,148 had previously been conducted. Here we present new radio observations
resulting in a detection of the system with both the VLA and the GMRT.

\begin{figure}
\centering
\epsfig{file=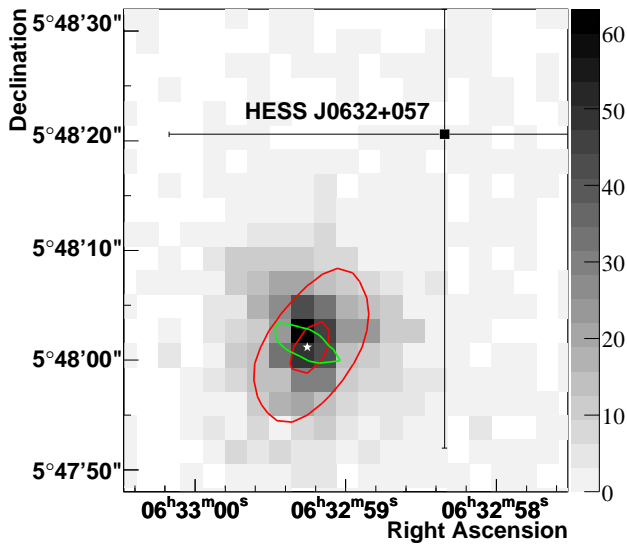, width=1\linewidth, clip=} \\
\caption{XMM-Newton count map \citep[see][for details]{0632:XMM} overlaid with
VLA 5\,GHz contours (0.2 and 0.4 mJy/beam; red)
and a single GMRT 1.28\,GHz contour (0.2 mJy/beam; green). Beam sizes as in Fig. 1. The position of MWC\,148 is
shown with a white star. The HESS best-fit position (black square) and 1$\sigma$ errors are also shown.}
\label{fig:zoom}
\end{figure}

\section{The new radio data}

\subsection{VLA data}

HESS\,J0632+057 was observed for 6 hours at 5\,GHz with the Very Large Array\footnote{The National Radio
Astronomy Observatory is a facility of the National Science Foundation
operated under cooperative agreement by Associated Universities, Inc.}
(VLA) in D-configuration 
(angular resolution $\approx$14$''$) during July -- September 2008 (program AS944). The time was divided into
three 2\,hour observations separated by $\sim$1 month. The data were calibrated using the NRAO AIPS \citep[Astronomical Image Processing System;][]{AIPS} software package and then loaded into
\emph{DIFMAP}\citep{Difmap} for additional editing and imaging. The flux density scale was set using a scan of 3C\,147
at the end of each observation and the phase was monitored with scans of the calibration source 0632+103. 
The off-source rms in each observation (0.03 -- 0.04\,mJy/beam) was estimated from large
sourceless boxes, far from the phase centre and was found to be close to the
thermal noise limit for a 2 hour observation of $\approx$0.03\,mJy/beam.

Five unresolved sources were found above 5$\sigma$ within the 9$'$ primary beam radius of the best fit position of HESS\,J0632+057.
\textbf{One of these sources (hereafter source \#3) is located at 06$^{\rm h}$32$^{\rm m}$59.24$^{\rm s}$ $\pm$ 0.3$''$, +05$^{\circ}$48'00.8'' $\pm$ 0.3$''$, consistent with the position of MWC\,148. This position was derived from a two-dimensional Gaussian fit to the brightest single VLA observation. The error on this measurement is statistical only.} Emission is detected
from this source in all three observations and the measured flux
varies significantly from observation to observation: from $0.19\pm0.04$\,mJy to
$0.41\pm0.04$\,mJy ($\chi^2$/dof for a constant fit = 19.4/2, chance probability = 6.1$\times$10$^{-5}$), see Fig.~3.
Additionally, since each 2\,hour observation consisted of four 21\,minute scans of the region surrounding HESS\,J0632+057, we searched for shorter timescale (intrahour) variability. 
%Scan-by-scan
%lightcurves were created to investigate variability on shorter timescales.
No evidence was found for short-term ($\sim$1\,hour) variability of source \#3 in the
scan-by-scan light-curves.
Source \#3 was modelled with an elliptical 2D Gaussian in the map plane resulting in
a size of 7$''$ by 4$''$ (1$\sigma$), consistent with an unresolved source.
%compatible with the beam size for this observation. 
%Thus the measured size should be considered as an upper limit
%on the intrinsic size.
%%JAH> comes later

A further
3\,hour observation at 5 and 8.5\,GHz with the VLA in the high resolution A-configuration (program AS967)
was taken in October 2008.
%%% with the aim of constraining the size of the emission region of this object.
However, source \#3 was not detected during this observation, presumably due to a low flux state
of this object. 
A plausible alternative reason for the non-detection
is that the source is extended on scales significantly larger than the $0.4''$ beam (but smaller than 
the 2$''$ 1.28\,GHz GMRT beam), reducing the signal/noise achievable in this configuration.
To circumvent this issue, the A configuration data were tapered to match the 2$''$ 1.28\,GHz GMRT beam and rms point source limits at 5 and 8\,GHz
were measured and plotted in Fig.~3. All four of the field sources detected
during the D configuration observations were also present in the A configuration
observations.
%, with consistent fluxes.

\textbf{The next closest source to MWC\,148 (source \#2) lies on the edge of the rms size limit of the TeV emission, at 06$^{\rm h}$33$^{\rm m}$06.01$^{\rm s}$ $\pm$ 0.3$''$, $+$05$^{\circ}$47$'$49.63$''$$\pm$ 0.3$''$. The source displays a power-law spectrum with $\alpha = 0.9 \pm 0.1$ and has a flux density at 5\,GHz of 0.7 $\pm$ 0.2\,mJy/beam. A multi-wavelength catalogue search resulted in no counterparts at this position.}
One of the field sources (source \#1) shows significant variability both on the
$\sim$month separation timescale of these observations and on shorter ($\sim$hour) timescales.
This source is located at 06h33m06.59s $\pm$ 0.3$''$,  +05d53$'$39$''$24 $\pm$ 0.3$''$, and is visible both in the NVSS archive \citep[1.4\,GHz,][]{Condon98}
with a flux of $\sim$15\,mJy (cf $\sim$12\,mJy in the GMRT 1.28\,GHz observations) and in the XMM-Newton image
of this field. Constant fluxes were recorded for all remaining field sources in all observations.

\subsection{GMRT data}

We also observed the field of HESS\,J0632+057, using the Giant Metrewave Radio Telescope\footnote{GMRT is run by the National Centre for Radio Astrophysics of
the Tata Institute of Fundamental Research.} (GMRT) \citep{Swarup91} 
during June -- September 2008.
The GMRT array has 30 antennas, arranged in roughly a `Y' configuration over 25 km area. The target field was
observed for a total of 36 hours with 6+6 slots of 3 hours each at 1280 and 610/235 MHz.
We
observed simultaneously at 235 and 610 MHz, using synthesised bandwidths of 6 
%(80 channels)
and 16 
%(126 channels) 
MHz, respectively. The 1280 MHz observation was carried out 
%in the default observing mode 
with
a bandwidth of 16 MHz 
%(126 channels) 
in each of the two available sidebands. Each frequency channel is 125 kHz
in width, enabling the removal
of narrow-band radio frequency interference (RFI). The sources 3C\,147 and 3C\,48 were
observed at the beginning and end of the observations and used as amplitude and bandpass calibrators to set the
flux density scale. The sources 0632+103 and 0521+166 were used as phase calibrators at 1280 and 610/235 MHz
respectively. All frequency channels affected by strong narrow-band RFI were flagged and removed from the entire data set
using the AIPS software package.

A faint, unresolved radio source is detected at the position of MWC\,148 (see Fig.~1)
in five of the six 1280\,MHz observations, with an
average flux of 0.68\,$\pm$\,0.04\,mJy (see Fig.~3).
\textbf{The best fit position of this source from the full data set is 06$^{\rm h}$32$^{\rm m}$59.29$^{\rm s}$\,$\pm$\,0.5$''$, +05$^{\circ}$48'01.2''\,$\pm$\,0.5$''$}.
There is no significant evidence for variability
in these data over the week to month timescales of the observations.
% The high angular resolution of the GMRT has helped to reveal that there is no substructure around 
% the HESS source and has placed an upper limit on the size of the emission region of $\sim$2$''$.
% JAH - we already said it is unresolved - and discussion comes later
MWC\,148 was not detected above the 3$\sigma$ level 
in the lower frequency observations.
Upper limits
on the flux at 610\,MHz and 235\,MHz have been calculated and included on the SED shown
in Fig.~4. 
The typical noise levels at 610 and 235 MHz were 0.74 and 1.82 mJy/beam respectively, the $<$1\,GHz data were more strongly affected by RFI, resulting in a substantial increase in the rms noise.

\begin{figure*}
\centering
\epsfig{file=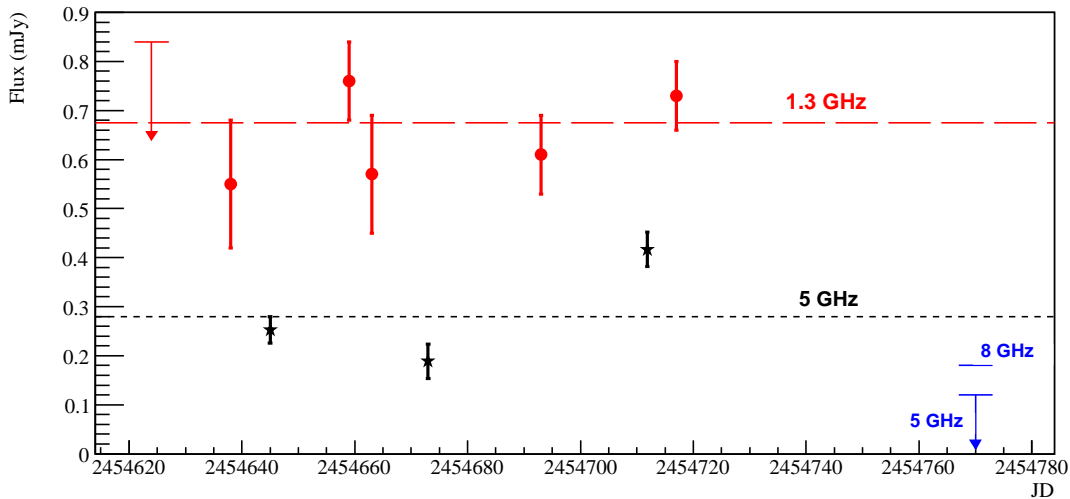, width=0.9\linewidth, clip=} \\
\caption{The radio flux density of MWC\,148 as a function of time (note that JD 2454620 is the 2$^{\rm nd}$ June 2008). The black \textbf{star-shaped} points denote observations taken 
	with the VLA and the red \textbf{circular} points are from the GMRT. The dotted lines represent the weighted 
	mean flux at each frequency. Flux density upper limits (at 99.8\% confidence level) from the VLA A configuration observations are
	plotted in blue.}
\label{fig:lc}
\end{figure*}

\section{Discussion}

\begin{table*}
\centering

{\footnotesize \begin{tabular}{l c c c c c c c c}

\hline\hline
Name & $D$ (kpc) & $L_{\rm r}$ & $L_{\rm X}$ & $L_{\rm GeV}$$^{\ast}$ & $L_{\rm TeV}$ & $\alpha$$_r$ & $\alpha$$_{X}$ & $\alpha$$_{\gamma}$ \\
% & (kpc) & (10$^{33}$erg\,s$^{-1}$) & (10$^{31}$erg\,s$^{-1}$) & (10$^{33}$erg\,s$^{-1}$) & (10$^{33}$erg\,s$^{-1}$) & & &\\
\hline
LS 5039 & 2.5 &1.3 & 5--50  & 70 & 4--11 & 0.46$^{\oplus}$ & 0.45 -- 0.6$^{\oslash}$ & 0.9 -- 1.5$^{\bowtie}$\\
LS I +61 303 & 2.0 & 1--17 & 3--9 & 60 & 8 & -0.6 -- 0.45$^{\dagger}$ & 0.53$^{\ddagger}$  & 1.6\,$\pm$\,0.2$^{\sharp}$\\
PSR B1259-63 & 1.5 & 0.02--0.3 & 0.3--6 & ...$^{!}$ & 2.3 & -2.2 -- 0.3$^{\amalg}$  & 0.78$^{\bullet}$  & 1.7\,$\pm$\,0.2$^+$\\
Cygnus X-1 & 2.2 & 0.3 & 10$^4$ & ...$^{!}$ & 12 & 0.1$^{\diamond}$  & 0.8$^{\star}$  & 2.2\,$\pm$\,0.6\\
HESS\,J0632+057 & 1.5  & 0.003 & 0.13$^{\bigtriangleup}$ & $<$9 & 0.9$^{\otimes}$ & 0.6 & 0.26$^{\bigtriangleup}$ & 1.5\,$\pm$\,0.3$^{\otimes}$\\
\hline
     \end{tabular}
      \caption{Properties of the TeV emitting binaries, adapted from
\citet{Paredes08}. Note that the spectral indices are defined by
$F_{\nu} \propto \nu ^{-\alpha}$ or equivalently $dN/dE \propto
E^{-(1+\alpha)}$. All luminosities are in units of
10$^{33}$erg\,s$^{-1}$ except the radio luminosities which are in
units of 10$^{31}$erg\,s$^{-1}$. Luminosities are given for the
following ranges: L$_{\rm r}$: 0.1 --100\,GHz, L$_{\rm X}$: 1 -- 10
keV, L$_{\rm GeV}$: 1 -- 10 GeV, L$_{\rm TeV}$: 0.2 -- 10
TeV. $^{\oplus}$\citet{Marti98}, $^{\oslash}$\citet{Takahashi09},
$^{\bowtie}$\citet{HESS:5039:06}, $^{\dagger}$\citet{Gregory02},
$^{\ddagger}$\citet{MAGIC:LSI}, $^{\sharp}$\citet{MAGIC:LSI:09},
$^{\amalg}$\citet{Johnston05}, $^{\bullet}$\citet{Esposito07},
$^+$\citet{HESS:1259}, $^{\diamond}$\citet{Pandey07},
$^{\star}$\citet{Miller05}, $^{\bigtriangleup}$\citet{0632:XMM},
$^{\otimes}$\citet{0632:HESS}. $^{\ast}$GeV luminosity measurements are
from the Fermi Bright Source List \citep{Fermi:cat0}, with upper
limits estimated from non-detection at a 10$\sigma$ level after three
months of observations. Note that the association of GeV emission with
LS\,5039 is based only on positional coincidence. $^{!}$No GeV
detection reported yet.} 
}
  \end{table*}

% Tycho position: 06 32 59.236	+05 48 01.04

The radio properties of source \#3 can be summarised as follows.
%position and size
The best fit positions of source \#3 measured with the VLA and GMRT are
consistent both with each other and with the position of MWC\,148
\citep[06$^{\rm h}$32$^{\rm m}$59.236$^{\rm s}$, +05$^{\circ}$48'01.04'';][]{Hog00}.
The new GMRT data provides the best constraint on the size of the radio emitting region
in MWC\,148. The upper limit on the source size (along the short axis of the GMRT beam) is
2$''$. Assuming a distance to the star of 1.5 kpc, this corresponds to an upper limit
on the emission region of 3000\,AU.
%variability
Significant variability was detected in the 5\,GHz emission from MWC\,148 at the 5$\sigma$ level.
Variability at the same level may be present in the 1.28\,GHz data but not detectable above the noise. 
The non-detection during the VLA A configuration observations
further demonstrates the variable nature of the source.
The variability timescale must be longer than the $\approx$2\,hour scans
and shorter than the $\sim$month timescale separation of the observations.
%Spectral index. 
An estimate of the spectral index, based on the
average flux at 1.28\,GHz and 5\,GHz, was calculated for source \#3
to be $\alpha_r$ = 0.6 $\pm$ 0.2. However, this value should be treated with
caution due to the non-simultaneous nature of the 1.28\,GHz and 5\,GHz observations.

%positional associations
The new radio source lies less than $\sim$1$''$ from the
centroid of the X-ray emission \citep{0632:XMM}, and within the 1$\sigma$ error box of HESS\,J0632+057 \textbf{(see Fig. 2)}.
Thus we confidently identify this emission as the radio counterpart to MWC\,148.
It seems very likely that all these objects are associated and therefore we
can create an SED as shown in Fig.~4. The recent publication of the Fermi Bright Source List \citep{Fermi:cat0}
allows us to place constraints on the GeV emission from HESS\,J0632+057. As no source
was detected (above 10$\sigma$ in the first 3 months of operation) at the
position of HESS\,J0632+057, the EGRET source 3EG\,J0634+0521 can be ruled out as a potential
counterpart. Limits for the GeV emission after 3 months, and 1 year of Fermi data
have been added to the SED. 

The detection of TeV emission implies particle acceleration is taking place in
the source. The most natural explanation for the observed radio emission is then
optically thin synchrotron emission from accelerated electrons.
Synchrotron radiation with a spectral index $\sim$0.5 implies an underlying electron spectrum with $dN/dE \propto E^{2}$.
This can be interpreted as the injection spectrum.
% but it is difficult to avoid rapid cooling of even low energy electrons in such a system.
 IC cooling in the Thomson regime is hard to avoid for these electrons.
 If the observed
spectrum is cooled, then this implies a very hard injection spectrum.
This scenario was the one presented in \cite{0632:XMM}, in which 
there is a low energy cut-off in the injection spectrum, resulting
in an effectively mono-energetic injection for electrons below this 
cut-off energy. Synchrotron cooling of such an injection will produce a time averaged electron spectral index of 2 and an emission spectrum with $\alpha$=0.5. The simple one-zone model presented in \cite{0632:XMM}
and adjusted to fit the X-ray and TeV data,
results in a radio spectrum remarkably similar to that observed.
A better fit to the radio data can be achieved by adjusting the low-energy
cut-off or a change in injection index from 2.0 to 1.9 (see Fig.~4).
%MODEL --$>$
Assuming a one-zone model with the parameters described in \cite{0632:XMM},
$B$~=~70\,mG, $U_{\rm rad}$~=~1\,erg\,cm$^{-3}$, the radio emission
at 5\,GHz would be produced by $\sim$90~MeV electrons having a characteristic cooling time
due to IC losses of $\approx$2 days. Similarly, 1.3\,GHz emission would be
produced by 50\,MeV electrons with a cooling time of $\sim$4 days.
The observed variability timescale, in common with that seen in X-rays, is
comparable with the expected cooling times, even if the radio
emission region is somewhat larger than the few AU assumed.
We note that if the emission region was considerably smaller than 
this both significant free-free absorption at 1\,GHz in the stellar 
wind and $\gamma$-$\gamma$ absorption at a few hundred GeV on stellar 
photons would likely become apparent at some orbital phases.
At this stage neither effect can be excluded. Strictly simultaneous
multi-frequency radio data would be required to check for radio absorption features.
%Future GeV data may help to establish whether $\gamma$-$\gamma$ cascading plays a significant
%role in this object.

%Magnetic Bp star
There are two main scenarios that could plausibly explain the observed emission from
HESS\,J0632+057; firstly that the star is isolated,
with a magnetically confined wind region and secondly that the star
is in a binary system with an unseen compact companion.
Although there are some interesting similarities between MWC\,148
and the archetypal isolated magnetic Bp star $\sigma$ Ori E, including
similar X-ray variability timescales \citep{Skinner08,0632:HESS},
there are also significant differences.
The X-ray emission from $\sigma$ Ori E can be well fit by a two-temperature
\emph{apec} model with a hot component of $\approx$2.6\,keV whereas the X-ray emission from
MWC\,148 can only be fit with such a model if the temperature of the hot
component is well above 10\,keV, due to the substantial flux
in the 5--10\,keV band. 
%%% Then why does a BB with these temperatures fit OK ? - see our XMM paper...
At radio wavelengths
$\sigma$ Ori E displays a flat spectrum ($\alpha \sim$ 0)between 2--20\,GHz,
whereas the emission from MWC\,148 appears to have a negative spectral index. 
The radio light-curve of  $\sigma$ Ori E is strongly modulated with the orbital
period. This may also be the case with MWC\,148, but further observations 
are clearly needed to characterise its radio variability.

As can be clearly seen in Tab.~1 there is a strong resemblance between the known
$\gamma$-ray binaries and HESS\,J0632+057. Although HESS\,J0632+057 is a much weaker
source in all wavelengths, the similar spectral properties
make a convincing argument that this object does indeed represent
a new $\gamma$-ray emitting binary system.
\textbf{The apparent variability in the TeV emission of
HESS\,J0632+057, as inferred by the VERITAS collaboration
\citep{VERITAS:0632}, is also consistent with the properties of the known $\gamma$-ray binary systems.}
All objects in Tab.~1 exhibit a hard ($\alpha_{X}<$ 1) spectrum in the X-ray domain, with
HESS\,J0632+057 having a particularly hard spectrum.
The TeV spectral indices of all objects are consistent with each other and all considerably softer 
than the corresponding X-ray spectrum.
%%%possibly implying a similar emission mechanism.
All objects appear to have their peak energy output in the MeV to GeV range. Note that GeV emission
has only been detected using Fermi coincident with LS\,I\,+61\,303 and LS\,5039 \citep{Fermi:cat0}. 
Upper limits for the other sources have been estimated using detected sources close to the objects of interest
to determine the local detection threshold.
The $\gamma$-ray binaries all display variable radio emission with a wide range of spectral
properties including low energy spectral turnovers \citep[see e.g.][]{Godambe08}.
%%%% What is the difference? -Jim
The clear outlier in this group is Cygnus\,X-1, for example, although the
radio and VHE $\gamma$-ray luminosities are similar, the X-ray luminosity of
Cygnus\,X-1 is orders of magnitude greater than the other objects.
Resolved radio emission has been detected in both LS\,I\,+61\,303 and LS\,5039 \citep{Paredes98,Ribo08}
with emission regions of the order of $\sim$10\,AU. 
%In these resolved objects a common emission region for radio
%and X-ray emission seems unlikely 
%and a one-zone model (as
%described here) is likely not detailed enough to accurately explain the observed emission.
Our current observations
are unable to probe such small scales and more data is required
in order to resolve or further constrain the size of the emitting region and move 
towards a better understanding of this source.
Whilst the detection of variable radio emission from MWC\,148
further strengthens the case that HESS\,J0632+057 is indeed a TeV binary,
the scenario that this object is an (extremely unusual) isolated massive star remains an intriguing possibility.

\begin{figure*}
\centering
\epsfig{file=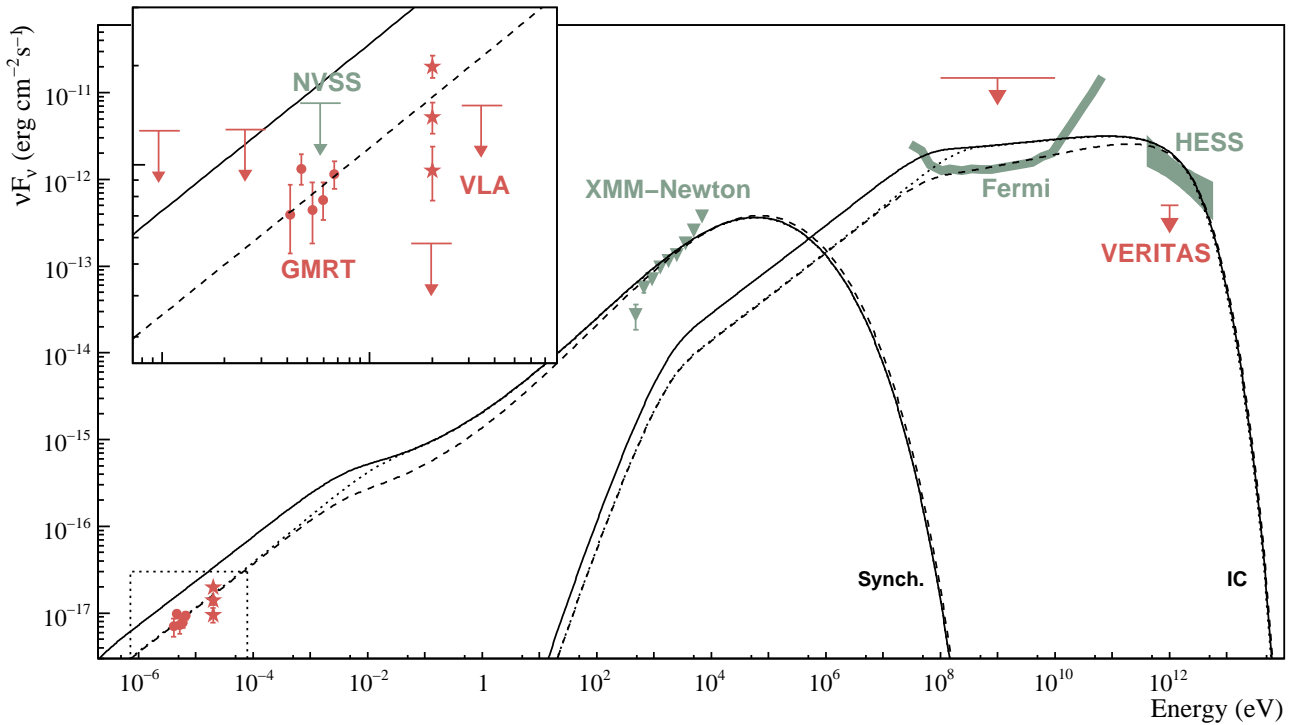, width=1\linewidth, clip=} \\
\caption{SED of HESS\,J0632+057 adapted from
\citep{0632:XMM}, with new data shown in red. 
The GMRT measurements have been plotted slightly offset from their observing 
frequency for clarity. The data are compared to
a one-zone model of non-thermal emission from electrons cooling in the
radiation and magnetic fields within a few AU of MWC\,148. The 
three model curves show an injection electron spectral index 2.0 (solid lines) and 1.9 (dashed lines), 
and an index of 2.0 but with a low energy cut-off at 2 GeV rather than 1 GeV (dotted line).
See \citet{0632:XMM} for more details. 
An upper limit for GeV emission from HESS\,J0632+057 based on three months of Fermi observations \citep{Fermi:cat0}
and the 1 year Fermi sensitivity curve are shown.
\textbf{An approximate energy flux limit from VERITAS is also shown \citep{VERITAS:0632}, highlighting the variable nature of the TeV emission.}
}
\label{fig:sed}
\end{figure*}

\section*{Acknowledgments}
We thank the staff of the GMRT who made these observations
possible. We thank Dr. S.~Roy and Prof.
V.~Kulkarni of NCRA, for the help provided during the GMRT observation and
data transmission to the Leiden Observatory, H.~E.~Wheelwright and
J.~J.~Stead for useful discussions and an anonymous referee for helpful comments.
J.A.H. is supported by a UK 
Science and Technology Facilities Council (STFC) Advanced Fellowship.
C.~C.~C.\ is supported by an appointment to the NASA Postdoctoral
Program at Goddard Space Flight Center, administered by Oak Ridge
Associated Universities through a contract with NASA. G.D. is supported by European Community contract ERC-StG-200911.

\footnotesize{
\bibliographystyle{aa}
\bibliography{Radio}{}

\begin{thebibliography}{35}
\expandafter\ifx\csname natexlab\endcsname\relax\def\natexlab#1{#1}\fi

\bibitem[{{Abdo} {et~al.}(2009){Abdo},{Aharonian}, {Akhperjanian},
  {Aye}, {Bazer-Bachi}, {Beilicke}, {Benbow}, {Berge}, {Berghaus},
  {Bernl{\"o}hr}, {Boisson}, {Bolz}, {Borrel}, {Braun}, {Breitling}, {Brown},
  {Gordo}, {Chadwick}, {Chounet}, {Cornils}, {Costamante}, \&
  {Degrange}}]{Fermi:cat0}
{Abdo}, A.~A., {Ackermann}, M., {Ajello}, M., {et~al.} 2009, ArXiv e-prints, 0902.1340

\bibitem[{{Acciari} {et~al.}(2009){Acciari}, {Aliu}, {Arlen}, {X}, {X}, \&
  {X}}]{VERITAS:0632}
{Acciari}, V.~A., {Aliu}, T., {Arlen}, M., {et~al.} 2009, ArXiv e-prints, 0905.3139

\bibitem[{{Aharonian} {et~al.}(2005{\natexlab{a}}){Aharonian}, {Akhperjanian},
  {Aye}, {Bazer-Bachi}, {Beilicke}, {Benbow}, {Berge}, {Berghaus},
  {Bernl{\"o}hr}, {Boisson}, {Bolz}, {Borrel}, {Braun}, {Breitling}, {Brown},
  {Gordo}, {Chadwick}, {Chounet}, {Cornils}, {Costamante}, {Degrange},
  {Dickinson}, {Djannati-Ata{\"i}}, {Drury}, {Dubus}, {Emmanoulopoulos},
  {Espigat}, {Feinstein}, {Fleury}, {Fontaine}, {Fuchs}, {Funk}, {Gallant},
  {Giebels}, {Gillessen}, {Glicenstein}, {Goret}, {Hadjichristidis}, {Hauser},
  {Heinzelmann}, {Henri}, {Hermann}, {Hinton}, {Hofmann}, {Holleran}, {Horns},
  {Jacholkowska}, {de Jager}, {Kh{\'e}lifi}, {Komin}, {Konopelko}, {Latham},
  {Le Gallou}, {Lemi{\`e}re}, {Lemoine-Goumard}, {Leroy}, {Lohse}, {Marcowith},
  {Martin}, {Martineau-Huynh}, {Masterson}, {McComb}, {de Naurois}, {Nolan},
  {Noutsos}, {Orford}, {Osborne}, {Ouchrif}, {Panter}, {Pelletier}, {Pita},
  {P{\"u}hlhofer}, {Punch}, {Raubenheimer}, {Raue}, {Raux}, {Rayner}, {Reimer},
  {Reimer}, {Ripken}, {Rob}, {Rolland}, {Rowell}, {Sahakian}, {Saug{\'e}},
  {Schlenker}, {Schlickeiser}, {Schuster}, {Schwanke}, {Siewert}, {Sol},
  {Spangler}, {Steenkamp}, {Stegmann}, {Tavernet}, {Terrier}, {Th{\'e}oret},
  {Tluczykont}, {Vasileiadis}, {Venter}, {Vincent}, {V{\"o}lk}, \&
  {Wagner}}]{HESS:5039}
{Aharonian}, F., {et~al.} (H.E.S.S. Collaboration)
  2005{\natexlab{a}}, Science, 309, 746

\bibitem[{{Aharonian} {et~al.}(2005{\natexlab{b}}){Aharonian}, {Akhperjanian},
  {Aye}, {Bazer-Bachi}, {Beilicke}, {Benbow}, {Berge}, {Berghaus},
  {Bernl{\"o}hr}, {Boisson}, {Bolz}, {Braun}, {Breitling}, {Brown}, {Bussons
  Gordo}, {Chadwick}, {Chounet}, {Cornils}, {Costamante}, {Degrange},
  {Djannati-Ata{\"i}}, {O'C.~Drury}, {Dubus}, {Emmanoulopoulos}, {Espigat},
  {Feinstein}, {Fleury}, {Fontaine}, {Fuchs}, {Funk}, {Gallant}, {Giebels},
  {Gillessen}, {Glicenstein}, {Goret}, {Hadjichristidis}, {Hauser},
  {Heinzelmann}, {Henri}, {Hermann}, {Hinton}, {Hofmann}, {Holleran}, {Horns},
  {de Jager}, {Johnston}, {Kh{\'e}lifi}, {Kirk}, {Komin}, {Konopelko},
  {Latham}, {Le Gallou}, {Lemi{\`e}re}, {Lemoine-Goumard}, {Leroy},
  {Martineau-Huynh}, {Lohse}, {Marcowith}, {Masterson}, {McComb}, {de Naurois},
  {Nolan}, {Noutsos}, {Orford}, {Osborne}, {Ouchrif}, {Panter}, {Pelletier},
  {Pita}, {P{\"u}hlhofer}, {Punch}, {Raubenheimer}, {Raue}, {Raux}, {Rayner},
  {Redondo}, {Reimer}, {Reimer}, {Ripken}, {Rob}, {Rolland}, {Rowell},
  {Sahakian}, {Saug{\'e}}, {Schlenker}, {Schlickeiser}, {Schuster}, {Schwanke},
  {Siewert}, {Skj{\ae}raasen}, {Sol}, {Steenkamp}, {Stegmann}, {Tavernet},
  {Terrier}, {Th{\'e}oret}, {Tluczykont}, {Vasileiadis}, {Venter}, {Vincent},
  {V{\"o}lk}, \& {Wagner}}]{HESS:1259}
{Aharonian}, F., {et~al.} (H.E.S.S. Collaboration)
  2005{\natexlab{b}}, \aap, 442, 1

\bibitem[{{Aharonian} {et~al.}(2006){Aharonian}, {Akhperjanian}, {Bazer-Bachi},
  {Beilicke}, {Benbow}, {Berge}, {Bernl{\"o}hr}, {Boisson}, {Bolz}, {Borrel},
  {Braun}, {Brown}, {B{\"u}hler}, {B{\"u}sching}, {Carrigan}, {Chadwick},
  {Chounet}, {Cornils}, {Costamante}, {Degrange}, {Dickinson},
  {Djannati-Ata{\"i}}, {O'C.~Drury}, {Dubus}, {Egberts}, {Emmanoulopoulos},
  {Espigat}, {Feinstein}, {Ferrero}, {Fiasson}, {Fontaine}, {Funk}, {Funk},
  {F{\"u}{\ss}ling}, {Gallant}, {Giebels}, {Glicenstein}, {Goret},
  {Hadjichristidis}, {Hauser}, {Hauser}, {Heinzelmann}, {Henri}, {Hermann},
  {Hinton}, {Hoffmann}, {Hofmann}, {Holleran}, {Horns}, {Jacholkowska}, {de
  Jager}, {Kendziorra}, {Kh{\'e}lifi}, {Komin}, {Konopelko}, {Kosack},
  {Latham}, {Le Gallou}, {Lemi{\`e}re}, {Lemoine-Goumard}, {Lohse}, {Martin},
  {Martineau-Huynh}, {Marcowith}, {Masterson}, {Maurin}, {McComb}, {Moulin},
  {de Naurois}, {Nedbal}, {Nolan}, {Noutsos}, {Orford}, {Osborne}, {Ouchrif},
  {Panter}, {Pelletier}, {Pita}, {P{\"u}hlhofer}, {Punch}, {Raubenheimer},
  {Raue}, {Rayner}, {Reimer}, {Reimer}, {Ripken}, {Rob}, {Rolland}, {Rowell},
  {Sahakian}, {Santangelo}, {Saug{\'e}}, {Schlenker}, {Schlickeiser},
  {Schr{\"o}der}, {Schwanke}, {Schwarzburg}, {Shalchi}, {Sol}, {Spangler},
  {Spanier}, {Steenkamp}, {Stegmann}, {Superina}, {Tavernet}, {Terrier},
  {Tluczykont}, {van Eldik}, {Vasileiadis}, {Venter}, {Vincent}, {V{\"o}lk},
  {Wagner}, \& {Ward}}]{HESS:5039:06}
{Aharonian}, F., {et~al.} (H.E.S.S. Collaboration) 2006,
  \aap, 460, 743

\bibitem[{{Aharonian} {et~al.}(2007){Aharonian}, {Akhperjanian}, {Bazer-Bachi},
  {Behera}, {Beilicke}, {Benbow}, {Berge}, {Bernl{\"o}hr}, {Boisson}, {Bolz},
  {Borrel}, {Braun}, {Brion}, {Brown}, {B{\"u}hler}, {B{\"u}sching},
  {Boutelier}, {Carrigan}, {Chadwick}, {Chounet}, {Coignet}, {Cornils},
  {Costamante}, {Degrange}, {Dickinson}, {Djannati-Ata{\"i}}, {Domainko},
  {O'C.~Drury}, {Dubus}, {Egberts}, {Emmanoulopoulos}, {Espigat}, {Farnier},
  {Feinstein}, {Fiasson}, {F{\"o}rster}, {Fontaine}, {Funk}, {Funk},
  {F{\"u}{\ss}ling}, {Gallant}, {Giebels}, {Glicenstein}, {Gl{\"u}ck}, {Goret},
  {Hadjichristidis}, {Hauser}, {Hauser}, {Heinzelmann}, {Henri}, {Hermann},
  {Hinton}, {Hoffmann}, {Hofmann}, {Holleran}, {Hoppe}, {Horns},
  {Jacholkowska}, {de Jager}, {Kendziorra}, {Kerschhaggl}, {Kh{\'e}lifi},
  {Komin}, {Kosack}, {Lamanna}, {Latham}, {Le Gallou}, {Lemi{\`e}re},
  {Lemoine-Goumard}, {Lohse}, {Martin}, {Martineau-Huynh}, {Marcowith},
  {Masterson}, {Maurin}, {McComb}, {Moulin}, {de Naurois}, {Nedbal}, {Nolan},
  {Noutsos}, {Olive}, {Orford}, {Osborne}, {Panter}, {Pedaletti}, {Pelletier},
  {Petrucci}, {Pita}, {P{\"u}hlhofer}, {Punch}, {Ranchon}, {Raubenheimer},
  {Raue}, {Rayner}, {Reimer}, {Ripken}, {Rob}, {Rolland}, {Rosier-Lees},
  {Rowell}, {Ruppel}, {Sahakian}, {Santangelo}, {Saug{\'e}}, {Schlenker},
  {Schlickeiser}, {Schr{\"o}der}, {Schwanke}, {Schwarzburg}, {Schwemmer},
  {Shalchi}, {Sol}, {Spangler}, {Steenkamp}, {Stegmann}, {Superina}, {Tam},
  {Tavernet}, {Terrier}, {Tluczykont}, {van Eldik}, {Vasileiadis}, {Venter},
  {Vialle}, {Vincent}, {V{\"o}lk}, {Wagner}, {Ward}, {Moriguchi}, \&
  {Fukui}}]{0632:HESS}
{Aharonian}, F., {et~al.} (H.E.S.S. Collaboration) 2007,
  \aap, 469, L1

\bibitem[{{Albert} {et~al.}(2009){Albert}, {Aliu}, {Anderhub}, {Antonelli},
  {Antoranz}, {Backes}, {Baixeras}, {Barrio}, {Bartko}, {Bastieri}, {Becker},
  {Bednarek}, {Berger}, {Bernardini}, {Bigongiari}, {Biland}, {Bock},
  {Bonnoli}, {Bordas}, {Bosch-Ramon}, {Bretz}, {Britvitch}, {Camara},
  {Carmona}, {Chilingarian}, {Commichau}, {Contreras}, {Cortina}, {Costado},
  {Covino}, {Curtef}, {Dazzi}, {DeAngelis}, {DeCea del Pozo}, {de los Reyes},
  {DeLotto}, {DeMaria}, {DeSabata}, {Delgado Mendez}, {Dominguez}, {Dorner},
  {Doro}, {Errando}, {Fagiolini}, {Ferenc}, {Fern{\'a}ndez}, {Firpo},
  {Fonseca}, {Font}, {Galante}, {L{\'o}pez}, {Garczarczyk}, {Gaug}, {Goebel},
  {Hayashida}, {Herrero}, {H{\"o}hne}, {Hose}, {Hsu}, {Huber}, {Jogler},
  {Kranich}, {La Barbera}, {Laille}, {Leonardo}, {Lindfors}, {Lombardi},
  {Longo}, {L{\'o}pez}, {Lorenz}, {Majumdar}, {Maneva}, {Mankuzhiyil},
  {Mannheim}, {Maraschi}, {Mariotti}, {Mart{\'{\i}}nez}, {Mazin}, {Meucci},
  {Meyer}, {Miranda}, {Mirzoyan}, {Mizobuchi}, {Moles}, {Moralejo}, {Nieto},
  {Nilsson}, {Ninkovic}, {Otte}, {Oya}, {Panniello}, {Paoletti}, {Paredes},
  {Pasanen}, {Pascoli}, {Pauss}, {Pegna}, {Perez-Torres}, {Persic}, {Peruzzo},
  {Piccioli}, {Prada}, {Prandini}, {Puchades}, {Raymers}, {Rhode}, {Rib{\'o}},
  {Rico}, {Rissi}, {Robert}, {R{\"u}gamer}, {Saggion}, {Saito}, {Salvati},
  {Sanchez-Conde}, {Sartori}, {Satalecka}, {Scalzotto}, {Scapin}, {Schmitt},
  {Schweizer}, {Shayduk}, {Shinozaki}, {Shore}, {Sidro}, {Sierpowska-Bartosik},
  {Sillanp{\"a}{\"a}}, {Sobczynska}, {Spanier}, {Stamerra}, {Stark}, {Takalo},
  {Tavecchio}, {Temnikov}, {Tescaro}, {Teshima}, {Tluczykont}, {Torres},
  {Turini}, {Vankov}, {Venturini}, {Vitale}, {Wagner}, {Wittek}, {Zabalza},
  {Zandanel}, {Zanin}, \& {Zapatero}}]{MAGIC:LSI:09}
{Albert}, J., {Aliu}, E., {Anderhub}, H., {et~al.} 2009, \apj, 693, 303

\bibitem[{{Albert} {et~al.}(2006){Albert}, {Aliu}, {Anderhub}, {Antoranz},
  {Armada}, {Asensio}, {Baixeras}, {Barrio}, {Bartelt}, {Bartko}, {Bastieri},
  {Bavikadi}, {Bednarek}, {Berger}, {Bigongiari}, {Biland}, {Bisesi}, {Bock},
  {Bordas}, {Bosch-Ramon}, {Bretz}, {Britvitch}, {Camara}, {Carmona},
  {Chilingarian}, {Ciprini}, {Coarasa}, {Commichau}, {Contreras}, {Cortina},
  {Curtef}, {Danielyan}, {Dazzi}, {De Angelis}, {de los Reyes}, {De Lotto},
  {Domingo-Santamar{\'{\i}}a}, {Dorner}, {Doro}, {Errando}, {Fagiolini},
  {Ferenc}, {Fern{\'a}ndez}, {Firpo}, {Flix}, {Fonseca}, {Font}, {Fuchs},
  {Galante}, {Garczarczyk}, {Gaug}, {Giller}, {Goebel}, {Hakobyan},
  {Hayashida}, {Hengstebeck}, {H{\"o}hne}, {Hose}, {Hsu}, {Isar}, {Jacon},
  {Kalekin}, {Kosyra}, {Kranich}, {Laatiaoui}, {Laille}, {Lenisa}, {Liebing},
  {Lindfors}, {Lombardi}, {Longo}, {L{\'o}pez}, {L{\'o}pez}, {Lorenz},
  {Lucarelli}, {Majumdar}, {Maneva}, {Mannheim}, {Mansutti}, {Mariotti},
  {Mart{\'{\i}}nez}, {Mase}, {Mazin}, {Merck}, {Meucci}, {Meyer}, {Miranda},
  {Mirzoyan}, {Mizobuchi}, {Moralejo}, {Nilsson}, {O{\~n}a-Wilhelmi},
  {Ordu{\~n}a}, {Otte}, {Oya}, {Paneque}, {Paoletti}, {Paredes}, {Pasanen},
  {Pascoli}, {Pauss}, {Pavel}, {Pegna}, {Persic}, {Peruzzo}, {Piccioli},
  {Poller}, {Pooley}, {Prandini}, {Raymers}, {Rhode}, {Rib{\'o}}, {Rico},
  {Riegel}, {Rissi}, {Robert}, {Romero}, {R{\"u}gamer}, {Saggion},
  {S{\'a}nchez}, {Sartori}, {Scalzotto}, {Scapin}, {Schmitt}, {Schweizer},
  {Shayduk}, {Shinozaki}, {Shore}, {Sidro}, {Sillanp{\"a}{\"a}}, {Sobczynska},
  {Stamerra}, {Stark}, {Takalo}, {Temnikov}, {Tescaro}, {Teshima}, {Tonello},
  {Torres}, {Torres}, {Turini}, {Vankov}, {Vitale}, {Wagner}, {Wibig},
  {Wittek}, {Zanin}, \& {Zapatero}}]{MAGIC:LSI}
{Albert}, J., {Aliu}, E., {Anderhub}, H., {et~al.} 2006, Science, 312, 1771

\bibitem[{{Albert} {et~al.}(2007){Albert}, {Aliu}, {Anderhub}, {Antoranz},
  {Armada}, {Baixeras}, {Barrio}, {Bartko}, {Bastieri}, {Becker}, {Bednarek},
  {Berger}, {Bigongiari}, {Biland}, {Bock}, {Bordas}, {Bosch-Ramon}, {Bretz},
  {Britvitch}, {Camara}, {Carmona}, {Chilingarian}, {Coarasa}, {Commichau},
  {Contreras}, {Cortina}, {Costado}, {Curtef}, {Danielyan}, {Dazzi}, {De
  Angelis}, {Delgado}, {de los Reyes}, {De Lotto}, {Domingo-Santamar{\'{\i}}a},
  {Dorner}, {Doro}, {Errando}, {Fagiolini}, {Ferenc}, {Fern{\'a}ndez}, {Firpo},
  {Flix}, {Fonseca}, {Font}, {Fuchs}, {Galante}, {Garc{\'{\i}}a-L{\'o}pez},
  {Garczarczyk}, {Gaug}, {Giller}, {Goebel}, {Hakobyan}, {Hayashida},
  {Hengstebeck}, {Herrero}, {H{\"o}hne}, {Hose}, {Hsu}, {Jacon}, {Jogler},
  {Kosyra}, {Kranich}, {Kritzer}, {Laille}, {Lindfors}, {Lombardi}, {Longo},
  {L{\'o}pez}, {L{\'o}pez}, {Lorenz}, {Majumdar}, {Maneva}, {Mannheim},
  {Mansutti}, {Mariotti}, {Mart{\'{\i}}nez}, {Mazin}, {Merck}, {Meucci},
  {Meyer}, {Miranda}, {Mirzoyan}, {Mizobuchi}, {Moralejo}, {Nieto}, {Nilsson},
  {Ninkovic}, {O{\~n}a-Wilhelmi}, {Otte}, {Oya}, {Panniello}, {Paoletti},
  {Paredes}, {Pasanen}, {Pascoli}, {Pauss}, {Pegna}, {Persic}, {Peruzzo},
  {Piccioli}, {Prandini}, {Puchades}, {Raymers}, {Rhode}, {Rib{\'o}}, {Rico},
  {Rissi}, {Robert}, {R{\"u}gamer}, {Saggion}, {Saito}, {S{\'a}nchez},
  {Sartori}, {Scalzotto}, {Scapin}, {Schmitt}, {Schweizer}, {Shayduk},
  {Shinozaki}, {Shore}, {Sidro}, {Sillanp{\"a}{\"a}}, {Sobczynska}, {Stamerra},
  {Stark}, {Takalo}, {Temnikov}, {Tescaro}, {Teshima}, {Torres}, {Turini},
  {Vankov}, {Vitale}, {Wagner}, {Wibig}, {Wittek}, {Zandanel}, {Zanin}, \&
  {Zapatero}}]{MAGIC:cygnusX}
{Albert}, J., {Aliu}, E., {Anderhub}, H., {et~al.} 2007, \apjl, 665, L51

\bibitem[{{Borra} \& {Landstreet}(1979)}]{Borra79}
{Borra}, E.~F. \& {Landstreet}, J.~D. 1979, \apj, 228, 809

\bibitem[{{Bridle} \& {Greisen}(1994)}]{AIPS}
{Bridle}, A.~H. \& {Greisen}, E.~W. 1994, AIPS memo 87; Charlottesville: NRAO

\bibitem[{{Clark} {et~al.}(2001){Clark}, {Reig}, {Goodwin}, {Larionov}, {Blay},
  {Coe}, {Fabregat}, {Negueruela}, {Papadakis}, \& {Steele}}]{Clark01}
{Clark}, J.~S., {Reig}, P., {Goodwin}, S.~P., {et~al.} 2001, \aap, 376, 476

\bibitem[{{Condon} {et~al.}(1998){Condon}, {Cotton}, {Greisen}, {Yin},
  {Perley}, {Taylor}, \& {Broderick}}]{Condon98}
{Condon}, J.~J., {Cotton}, W.~D., {Greisen}, E.~W., {et~al.} 1998, \aj, 115,
  1693

\bibitem[{{Dubus}(2006)}]{Dubus06}
{Dubus}, G. 2006, \aap, 456, 801

\bibitem[{{Esposito} {et~al.}(2007){Esposito}, {Caraveo}, {Pellizzoni}, {de
  Luca}, {Gehrels}, \& {Marelli}}]{Esposito07}
{Esposito}, P., {Caraveo}, P.~A., {Pellizzoni}, A., {et~al.} 2007, \aap, 474,
  575

\bibitem[{{Godambe} {et~al.}(2008){Godambe}, {Bhattacharyya}, {Bhatt}, \&
  {Choudhury}}]{Godambe08}
{Godambe}, S., {Bhattacharyya}, S., {Bhatt}, N., \& {Choudhury}, M. 2008,
  \mnras, 390, L43

\bibitem[{{Gregory}(2002)}]{Gregory02}
{Gregory}, P.~C. 2002, \apj, 575, 427

\bibitem[{{Guti{\'e}rrez-Soto} {et~al.}(2007){Guti{\'e}rrez-Soto}, {Fabregat},
  {Suso}, {Lanzara}, {Garrido}, {Hubert}, \& {Floquet}}]{Gutierrez07}
{Guti{\'e}rrez-Soto}, J., {Fabregat}, J., {Suso}, J., {et~al.} 2007, \aap, 476,
  927

\bibitem[{{Hinton} {et~al.}(2009){Hinton}, {Skilton}, {Funk}, {Brucker},
  {Aharonian}, {Dubus}, {Fiasson}, {Gallant}, {Hofmann}, {Marcowith}, \&
  {Reimer}}]{0632:XMM}
{Hinton}, J.~A., {Skilton}, J.~L., {Funk}, S., {et~al.} 2009, \apjl, 690, L101

\bibitem[{{H{\o}g} {et~al.}(2000){H{\o}g}, {Fabricius}, {Makarov}, {Urban},
  {Corbin}, {Wycoff}, {Bastian}, {Schwekendiek}, \& {Wicenec}}]{Hog00}
{H{\o}g}, E., {Fabricius}, C., {Makarov}, V.~V., {et~al.} 2000, \aap, 355, L27

\bibitem[{{Johnston} {et~al.}(2005){Johnston}, {Ball}, {Wang}, \&
  {Manchester}}]{Johnston05}
{Johnston}, S., {Ball}, L., {Wang}, N., \& {Manchester}, R.~N. 2005, \mnras,
  358, 1069

\bibitem[{{Marti} {et~al.}(1998){Marti}, {Paredes}, \& {Ribo}}]{Marti98}
{Marti}, J., {Paredes}, J.~M., \& {Ribo}, M. 1998, \aap, 338, L71

\bibitem[{{Miller} {et~al.}(2005){Miller}, {Wojdowski}, {Schulz}, {Marshall},
  {Fabian}, {Remillard}, {Wijnands}, \& {Lewin}}]{Miller05}
{Miller}, J.~M., {Wojdowski}, P., {Schulz}, N.~S., {et~al.} 2005, \apj, 620,
  398

\bibitem[{{Morgan} {et~al.}(1955){Morgan}, {Code}, \& {Whitford}}]{Morgan55}
{Morgan}, W.~W., {Code}, A.~D., \& {Whitford}, A.~E. 1955, \apjs, 2, 41

\bibitem[{{Pandey} {et~al.}(2007){Pandey}, {Rao}, {Ishwara-Chandra},
  {Durouchoux}, \& {Manchanda}}]{Pandey07}
{Pandey}, M., {Rao}, A.~P., {Ishwara-Chandra}, C.~H., {Durouchoux}, P., \&
  {Manchanda}, R.~K. 2007, \aap, 463, 567

\bibitem[{{Paredes}(2008)}]{Paredes08}
{Paredes}, J.~M. 2008, in AIP Conference Series, Vol.
  1085, 157--168

\bibitem[{{Paredes} {et~al.}(1998){Paredes}, {Massi}, {Estalella}, \&
  {Peracaula}}]{Paredes98}
{Paredes}, J.~M., {Massi}, M., {Estalella}, R., \& {Peracaula}, M. 1998, \aap,
  335, 539

\bibitem[{{Phillips} \& {Lestrade}(1988)}]{Phillips88}
{Phillips}, R.~B. \& {Lestrade}, J.-F. 1988, \nat, 334, 329

\bibitem[{{Rib{\'o}} {et~al.}(2008){Rib{\'o}}, {Paredes}, {Mold{\'o}n},
  {Mart{\'{\i}}}, \& {Massi}}]{Ribo08}
{Rib{\'o}}, M., {Paredes}, J.~M., {Mold{\'o}n}, J., {Mart{\'{\i}}}, J., \&
  {Massi}, M. 2008, \aap, 481, 17

\bibitem[{{Shepherd} {et~al.}(1994){Shepherd}, {Pearson}, \& {Taylor}}]{Difmap}
{Shepherd}, M.~C., {Pearson}, T.~J., \& {Taylor}, G.~B. 1994, in BAAS, Vol.~26, 987--989

\bibitem[{{Skinner} {et~al.}(2008){Skinner}, {Sokal}, {Cohen}, {Gagn{\'e}},
  {Owocki}, \& {Townsend}}]{Skinner08}
{Skinner}, S.~L., {Sokal}, K.~R., {Cohen}, D.~H., {et~al.} 2008, \apj, 683, 796

\bibitem[{{Skrutskie} {et~al.}(2006){Skrutskie}, {Cutri}, {Stiening},
  {Weinberg}, {Schneider}, {Carpenter}, {Beichman}, {Capps}, {Chester},
  {Elias}, {Huchra}, {Liebert}, {Lonsdale}, {Monet}, {Price}, {Seitzer},
  {Jarrett}, {Kirkpatrick}, {Gizis}, {Howard}, {Evans}, {Fowler}, {Fullmer},
  {Hurt}, {Light}, {Kopan}, {Marsh}, {McCallon}, {Tam}, {Van Dyk}, \&
  {Wheelock}}]{2MASS:new}
{Skrutskie}, M.~F., {Cutri}, R.~M., {Stiening}, R., {et~al.} 2006, \aj, 131,
  1163

\bibitem[{{Swarup} {et~al.}(1991){Swarup}, {Ananthakrishnan}, {Kapahi}, {Rao},
  {Subrahmanya}, \& {Kulkarni}}]{Swarup91}
{Swarup}, G., {Ananthakrishnan}, S., {Kapahi}, V.~K., {et~al.} 1991, Current
  Science, 60, 95

\bibitem[{{Takahashi} {et~al.}(2009){Takahashi}, {Kishishita}, {Uchiyama},
  {Tanaka}, {Yamaoka}, {Khangulyan}, {Aharonian}, {Bosch-Ramon}, \&
  {Hinton}}]{Takahashi09}
{Takahashi}, T., {Kishishita}, T., {Uchiyama}, Y., {et~al.} 2009, \apj, 697,
  592

\bibitem[{{The} {et~al.}(1994){The}, {de Winter}, \& {Perez}}]{The94}
{The}, P.~S., {de Winter}, D., \& {Perez}, M.~R. 1994, \aaps, 104, 315

\bibitem[{{Townsend} {et~al.}(2007){Townsend}, {Owocki}, \&
  {Ud-Doula}}]{Townsend07}
{Townsend}, R.~H.~D., {Owocki}, S.~P., \& {Ud-Doula}, A. 2007, \mnras, 382, 139

\bibitem[{{Yudin} \& {Evans}(1998)}]{Yudin98}
{Yudin}, R.~V. \& {Evans}, A. 1998, \aaps, 131, 401

\end{thebibliography}
}

\end{document}